\documentclass[sigconf]{acmart}

\usepackage{subcaption}
\usepackage{xcolor}
\usepackage{soul} 
\soulregister\cite7
\soulregister\ref7
\usepackage{cleveref}

\AtBeginDocument{%
 \providecommand\BibTeX{{%
  \normalfont B\kern-0.5em{\scshape i\kern-0.25em b}\kern-0.8em\TeX}}}

\copyrightyear{2025}
\acmYear{2025}
\setcopyright{rightsretained}
\acmConference[CHI '25]{The CHI Conference on Human Factors in Computing Systems}{April 26-May 1, 2025}{Yokohama, Japan}
\acmBooktitle{The CHI Conference on Human Factors in Computing Systems (CHI '25), April 26-May 1, 2025, Yokohama, Japan}

\begin{document}

\title{The Goldilocks Time Window for Proactive Interventions in Wearable AI Systems}


\author{Cathy Mengying Fang}

\author{Wazeer Zulfikar}

\author{Pattie Maes}
\email{{catfang,wazeer,pattie}@media.mit.edu}

\affiliation{%
 \institution{MIT Media Lab}
 \city{Cambridge}
 \country{USA}
}

\author{Yasith Samaradivakara}

\author{Suranga Nanayakkara}
\email{{yasith,suranga}@ahlab.org}
\affiliation{%
 \institution{Augmented Human Lab, National University of Singapore}
 \city{Singapore}
 \country{Singapore}
}


\renewcommand{\shortauthors}{Fang, et al.}

\begin{abstract}
As AI systems become increasingly integrated into our daily lives and into wearable form factors, there's a fundamental tension between their potential to proactively assist us and the risk of creating intrusive, dependency-forming experiences. This work proposes the concept of a Goldilocks Time Window---a contextually adaptive time window for proactive AI systems to deliver effective interventions. We discuss the critical factors that determine the time window, and the need of a framework for designing and evaluating proactive AI systems that can navigate this tension successfully.
 
\end{abstract}

\maketitle

\section{Introduction}

Most systems for assistance and behavior change interact with users at a System 2 level—engaging individuals in slow, effortful thinking \cite{adams2015mindless}. We argue that in order for wearable AI systems to effectively engage users at a System 1 level—where thinking is fast and automatic \cite{kahneman2011thinking}—they need to be proactive. For instance, a person who wants to be healthy but subconsciously reaches out for a tasty yet unhealthy dessert would have liked an external intervention at that moment. ``Just-in-time adaptive interventions'' (JITAIs) solidified this concept by focusing on delivering the right type and amount of support at the right moment, adapting to an individual’s dynamic internal and contextual state \cite{nahum2018just, orzikulova2024time2stop}. 

Current proactive frameworks, however, mainly focus on conversational AI in text-based systems \cite{deng2025proactive,amershi2019guidelines, diebel2025ai}. Proactive interventions now take the form of various modalities such as audio, visual, haptic and more. Wearable multi-modal devices, such as smart glasses and head-mounted displays, allow the system to reference data in real-time and possibility have more avenues of intervening \cite{arakawa2024, Wang2013,lee2024gazepointar}. Thus, wearable AI can detect and respond to behavioral patterns in real-time, offering timely interventions when users have not consciously decided to seek external support but would still benefit from it. Building on these advancements, it is crucial to examine not only the capabilities of proactive wearable AI interventions but also how they can optimize intervention timing to maximize their effectiveness in real-world contexts.

This paper explores the critical timing dimension of proactive AI interventions, introducing the concept of the "Goldilocks Time Window"—the optimal window of time when interventions are neither too early (risking false positives and unnecessary interruptions) nor too late (when effectiveness diminishes). Specifically, we identify several key factors that determine effective intervention timing, such as contextual environment, delivery modality, and social awareness. We argue that successful proactive AI must balance predictability with strategic unexpectedness and  maintain alignment with user intent while preserving agency. We call for a standardized method for evaluating the timing dimension of proactive systems and assess intervention effectiveness, offering designers and researchers a structured approach to creating AI systems that know not just what to say, but precisely when to say it.

\begin{figure}[b!]
\vspace{-10px}
\includegraphics[width=.5\linewidth]{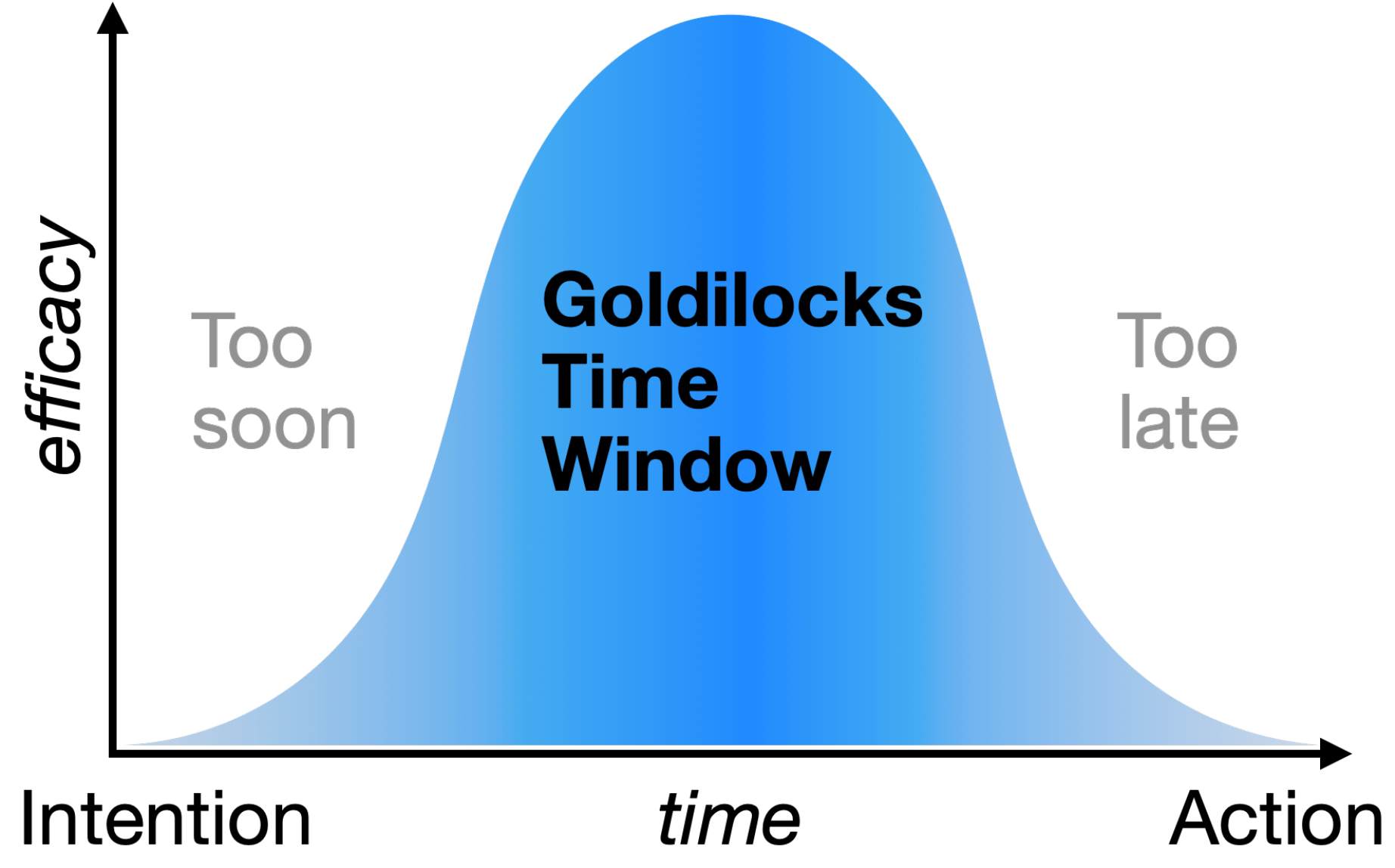}
\caption{The Goldilocks Time Window}
\label{fig:concept}
\vspace{-15px}
\end{figure}

\section{The Goldilocks Time Window}

With the introduction of foundation models like Large Language Models, interactions between users and systems are no longer constrained by graphical user interfaces (GUI); the communication with these systems becomes more \textbf{temporal} and the interactions becomes implicit, where the systems infer user intent based on additional information \cite{morris2025hci}. This shift introduces a critical challenge: determining when AI systems should intervene without being explicitly prompted by the user. Imagine a helicopter parent constantly hovering over your shoulder—this metaphor captures the risk of poorly-timed proactive AI.

From the user's thought to action, there is an optimum time window for intervention, which we call the ``Goldilocks Time Window'' (\Cref{fig:concept}). It has paradoxical properties: the efficacy of the intervention reduces as the timing gets closer to the user taking an undesirable action, but if the intervention happens too early, the probability of a false positive rate increases and may case unnecessary interruptions. Consider fire alarms. In high-stakes situation, the fire alarm abodes the user's trust that it must sound during genuine emergencies. However, a false positive is irritating, and repeated false positives lose the user's trust akin to ``The Boy who Cried Wolf'' effect. 

Effective intervention system must align with the user's greater intention or higher-level goals. At the same time, we argue the intervention's efficacy depends on unexpectedness. For example, the system chimes in and reminds the user about their health goals when the user subconsciously reaches for a chocolate bar. Thus, the intervention's specific timing provide value through their unexpectedness. Current user experience frameworks emphasizes the importance of predictability of user interfaces in building trust with the user \cite{nielsen10usability}. Proactive interventions seemingly contradict but complement the predictability of user interfaces with its strategic unexpectedness. 

In addition, well-timed interventions enhance rather than diminish user agency. Poorly-timed or overly frequent interventions can create dependency, where users rely on the system rather than developing their own capabilities. For example, a language learning tool that automatically corrects every error without requiring user engagement. Over time, the user's agency diminishes as the technology shifts from empowering tool to essential prosthetic.


\section{Factors that Affect the Goldilocks Time Window}

This optimal timing is influenced by several interconnected factors that work together to determine when an intervention will be most effective. First, the system should have social sensitivity. An intervention appropriate when someone is alone at home might be completely inappropriate during a job interview or important meeting. The delivery modality---whether through visual alerts, sounds, or haptic feedback---influences optimal timing, as different channels have varying levels of intrusiveness and attention-capturing ability. 

The alignment between immediate actions and longer-term goals creates tension that timing must navigate. Effective interventions balance immediate needs (like stress relief) with long-term objectives (such as health improvement). This balance shifts depending on the magnitude of potential consequences - interventions for minor dietary choices warrant different timing considerations than those addressing potential safety hazards. User history creates important precedents that shape future intervention effectiveness. Previous experiences with similar interventions establish expectations and response patterns that vary even between seemingly similar situations.


\section{Examples}

Here we use three projects done by the authors as examples of where proactive responses play a significant role in helping the user to achieve their respective goals.

\noindent
Mirai\cite{fang2025mirai} is a wearable AI system with an integrated camera, real-time speech processing, and personalized voice-cloning to provide proactive and contextual nudges for positive behavior change (\Cref{fig:mirai} in \Cref{apdx:mirai}). Mirai continuously monitors and analyzes the user's environment to anticipate their intentions, generating contextually-appropriate responses delivered in the user's own cloned voice. Noteably, the system implements a ``debouncer'' mechanism (\Cref{mirai-debouncer}) and a proactive agent (\Cref{mirai-prompt}) to infer the right timing to intervene. The debouncer mechanism ensures that responses are triggered only when there is a significant context change or at controlled intervals during stable states. Upon detecting new information, the proactive agent evaluates whether the update reflects a significant change in user behavior or context. It also determines the appropriate timing for a response—deciding whether to respond immediately or delay the response. If a response is warranted, it incorporates the updated understanding into the reply.

\noindent
Memoro\cite{zulfikar2024memoro} is a wearable audio-based memory assistant. Memoro uses a large language model (LLM) to infer the user’s memory needs in a conversational context, semantically search memories, and present minimal suggestions. The assistant has two interaction modes: Query Mode for voicing queries and Queryless Mode for on-demand predictive assistance, without explicit query \Cref{fig:memoro}. The Queryless mode is facilitated by the user requesting the memory assistant to understand the ongoing flow of the conversation and infer their precise memory need. 

\noindent
AiSee\cite{boldu2020aisee} is an assistive wearable interface that helps people with visual impairment overcome daily challenges, such as grocery shopping. The prototype consisted of a bone-conduction headset with a camera capable of processing and extracting features such as text, logos, and labels from the captured images. While evaluating the device with participants, the researchers observed that providing excessive information that is not 100\% relevant for the recognition produced confusion, and made the users lose their concentration on the outcome. Thus, future work on how a proactive version of the AI system could infer user's needs and focus and deliver only the critical information at the right moment.

\section{Call to Action}

This paper presents a case for the importance of timing in interventions delivered by proactive, wearable AI systems. We define the term ``Goldilocks Time Window'' to illustrate that the maximum efficacy of a proactive intevention happens when the timing is not too soon nor too late, and we identified factors that influence the Goldilocks Time Window. We used the authors' research prototypes as illustrative examples. Current evaluations of wearable systems are primarily centered around wearability and privacy \cite{olwal2020wearable,zulfikar2024memic}. We propose that future systems that involve a proactive approach should evaluate the systems' efficacy at achieving the right time window. In addition, because these wearable systems often require communications among multiple components, latency is also often used as part of the systems' technical evaluation, which primarily focus on the engineering aspects of achieving low latency. We advocate for a type of latency evaluation that focuses on the timing of the intervention delivery with respect to human needs. This ensures that the success metrics has ecological validity and grounded in practical, safe, and ethical use \cite{morris2025hci}. This new benchmark might look like a mix of qualitative and quantitative evaluations where the user provides open-ended feedback, given the variety in individuals' prior experience and contexts.



\bibliographystyle{ACM-Reference-Format}
\bibliography{sample-base}

\appendix

\section{Mirai}\label{apdx:mirai}
\subsection{Proactive prompt snippet} \label{mirai-prompt}
These are some contextual rules:
If you receive new information tagged as [NEW INFO] from "system", first consider whether this information indicates a new behavior of the user, and only if so you should initate a response.
If you are asked to remind the user, determine the right moment to say the reminder, do so succinctly under 5 words without causing much disruption.
A good rule of thumb for the right time to intervene is when the user hasn't been speaking for a while AND when you don't see someone else talking.

The full prompt can be found in \cite{fang2025mirai}.

\subsection{Debouncer overview} \label{mirai-debouncer}

The debouncer mechanism is mathematically represented as follows:

\[
U_t =
\begin{cases} 
1 & \text{if } S_t = "YES" \text{ and } S_t \neq S_{t-1}, \\
1 & \text{if } S_t = "YES" \text{ and } R_t \mod 3 = 0, \\
0 & \text{otherwise.}
\end{cases}
\]

Here, \( U_t \) determines whether a response is triggered at time \( t \). The classifier's state is represented by \( S_t \), with \( S_{t-1} \) denoting the previous state, and \( R_t \) representing the time step. A response is triggered under two conditions:
\begin{enumerate}
  \item The current state \( S_t \) is \texttt{"YES"} and differs from the previous state (\( S_t \neq S_{t-1} \)), indicating a significant context change.
  \item The current state \( S_t \) is \texttt{"YES"} and the time step \( R_t \) satisfies \( R_t \mod 3 = 0 \), which prevents repeated interruptions during stable states.
\end{enumerate}

This mechanism ensures that responses are contextually relevant, avoids excessive interruptions, and maintains responsiveness to meaningful changes.
\begin{figure}[H]
\includegraphics[width=1\linewidth]{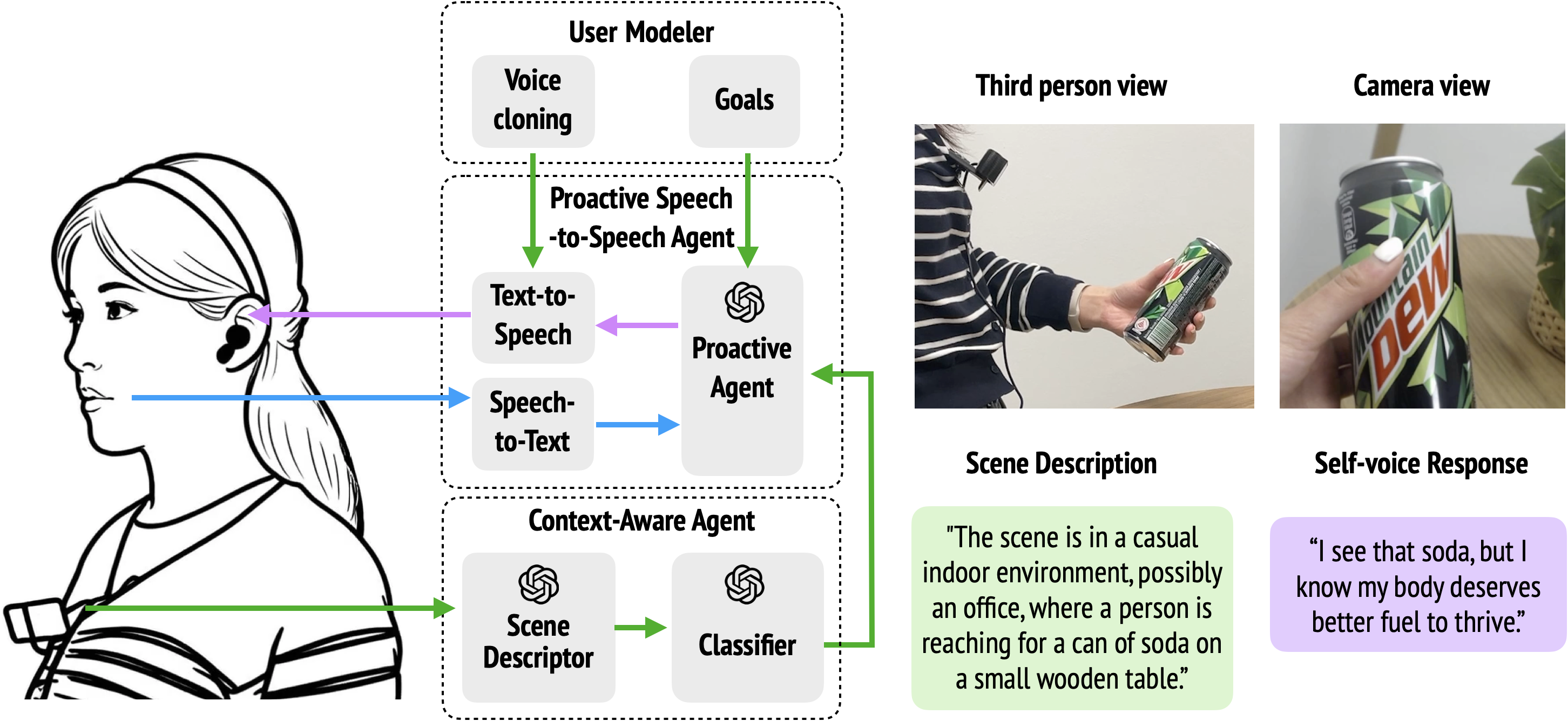}
\caption{Overview of Mirai.}
\label{fig:mirai}
\end{figure}

\section{Memoro}\label{apdx:memoro}

\subsection{Queryless prompt}

You are an assistant interface between user and a memory system. The user is engaged in a conversation with another human, and asks you in the middle for assistance. The assistant frames a query that the user would like to ask the memory system next at the end of the conversation. The recent conversation between the two humans is related to the relevant memories. The answer that the user would like to retrieve would not be in the recent conversation. The query should be very relevant to the end of the last sentence of the recent conversation. 

Recent conversation: <Current Context>

What do you think that the user would like to ask the memory system to finish or clarify his last sentence?

Query: [Generated Query]
\begin{figure}[H]
  \centering
  \includegraphics[width=1\linewidth]{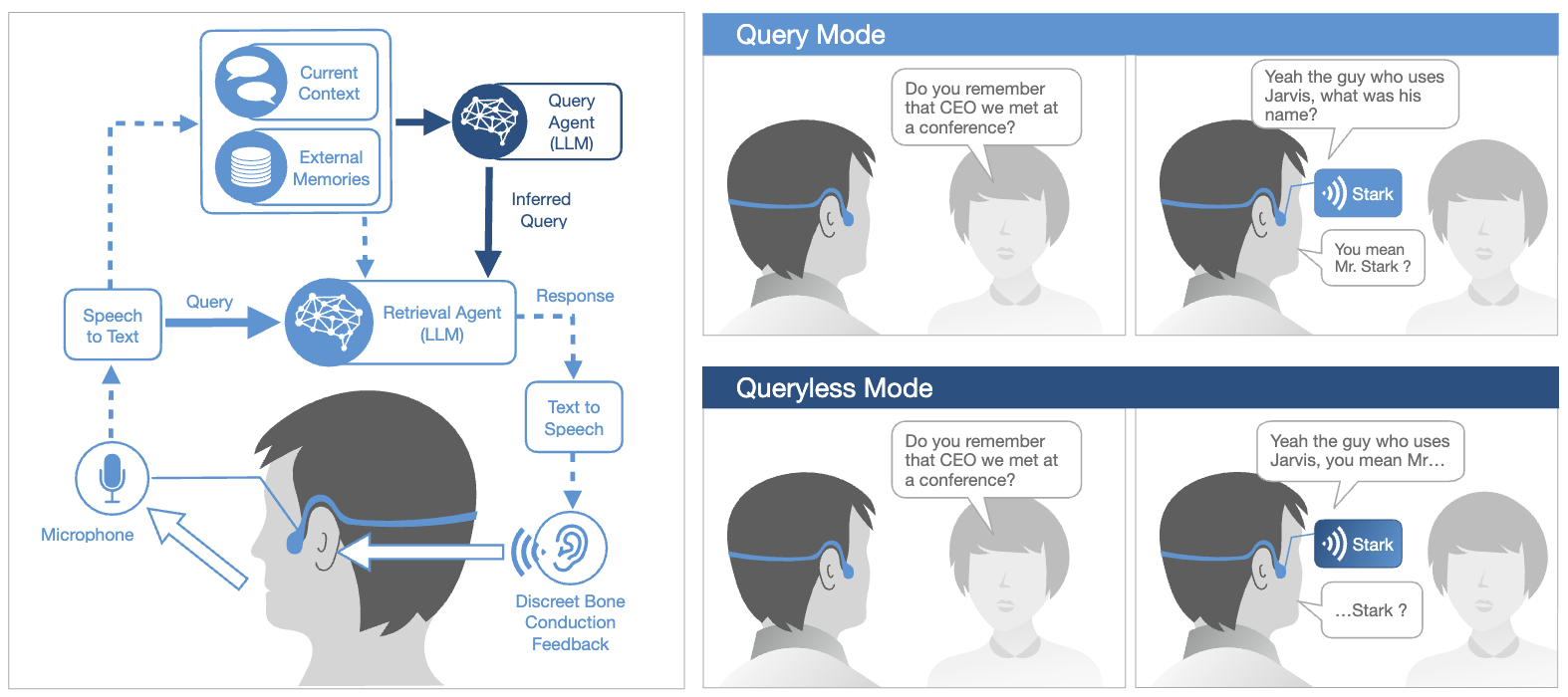}
  \caption{Memoro's interaction modes: (1) Query Mode where the user can ask contextual questions (2) Queryless Mode where the user can request predictive assistance and skip query formation. In both modes, responses are discreetly played back to the user using a bone conduction headset.}
  \label{fig:memoro}
\end{figure}

\section{AiSee}\label{apdx:aisee}

\begin{figure}[H]
  \centering
  \includegraphics[width=1\linewidth]{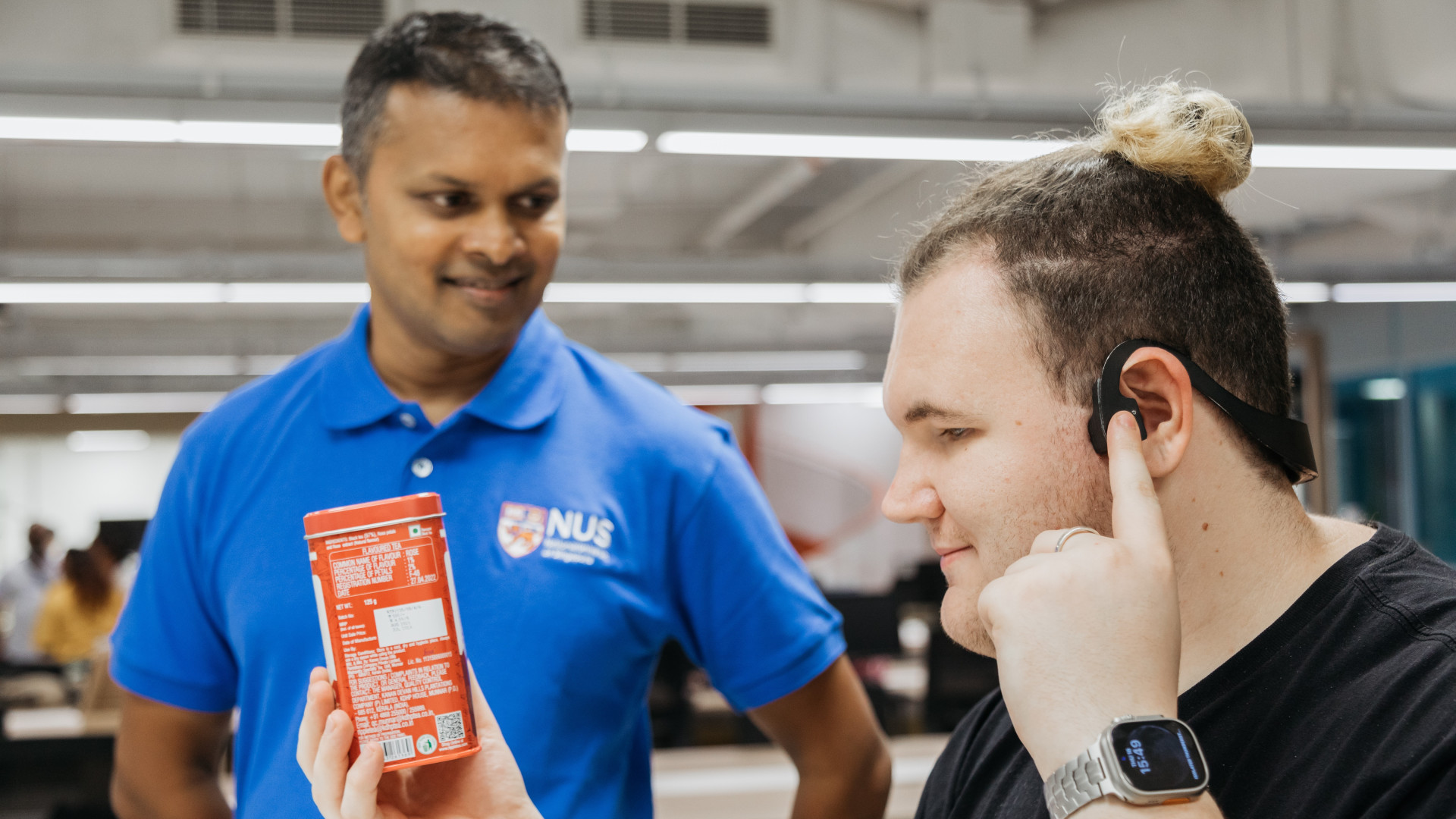}
  \caption{An assistive bone-conduction headset for people with visual impairment.}
  \label{fig:aisee}
\end{figure}
\end{document}